# Spatial distribution of core monomers in acrylamide-based core-shell microgels with linear swelling behaviour

Marian Cors[1, 2], Oliver Wrede[1], Lars Wiehemeier[1], Artem Feoktystov[3], Fabrice Cousin[4], Thomas Hellweg[1,5*], Julian Oberdisse[2*]

[1] Department of Physical and Biophysical Chemistry, Bielefeld University, Universitätsstr. 25, 33615 Bielefeld, Germany
[2] Laboratoire Charles Coulomb (L2C), University of Montpellier, CNRS, 34095 Montpellier, France.
[3] Forschungszentrum Jülich GmbH, Jülich Centre for Neutron Science JCNS at Heinz Maier-Leibnitz Zentrum MLZ, 85748 Garching, Germany.
[4] Laboratoire Léon Brillouin, UMR 12 CEA/CNRS, CEA Saclay, 91191 Gif Sur Yvette, France
[5] Lund Institute of Advanced Neutron and X-ray Science (LINXS), IDEON Building: Delta 5, Scheelevägen 19, 22370 Lund, Sweden

* Authors for correspondence : thomas.hellweg@uni-bielefeld.de, julian.oberdisse@umontpellier.fr

## Abstract

The peculiar linear temperature-dependent swelling of core-shell microgels has been conjectured to be linked to the core-shell architecture combining materials of different transition temperatures. Here the structure of pNIPMAM-core and pNNPAM-shell microgels in water is studied as a function of temperature using small-angle neutron scattering with selective deuteration. Photon correlation spectroscopy is used to scrutinize the swelling behaviour of the colloidal particles and reveals linear swelling. Moreover, these experiments are also employed to check the influence of deuteration on swelling. Using a form-free multi-shell reverse Monte Carlo approach, the small-angle scattering data are converted into radial monomer density profiles. The comparison of 'core-only' particles consisting of identical cores to fully hydrogenated core-shell microgels, and finally to H-core/D-shell architectures unambiguously shows that core and shell monomers display gradient profiles with strong interpenetration, leading to cores embedded in shells which are bigger than their isolated 'core-only' precursor particles. This surprising result is further generalized to different core cross-linker contents, for temperature ranges encompassing both transitions. Our analysis demonstrates that the internal structure of pNIPMAM-core and pNNPAM-shell microgels is heterogeneous and strongly interpenetrated, presumably allowing only progressive core swelling at temperatures intermediate to both transition temperatures, thus promoting linear swelling behaviour.

## Introduction

Acrylamide-based microgels are stable colloidal particles exhibiting a change in size upon modification of an external stimulus, in particular the temperature.[1-6] The most prominent examples are poly(*N*-isopropylacrylamide) (pNIPAM)-based microgels with a so-called volume phase transition temperature (VPTT) of ca. 33 °C, but other studies investigate microgels made of poly(*N*-isopropylmethacrylamide) (pNIPMAM, VPTT ≈ 45 °C) and poly(*N*-*n*-propylacrylamide) (pNNPAM, VPTT ≈ 23 °C), poly(N-vinylcaprolactam) (PVCL, VPTT≈ 32 °C).[1-5,7-19] In the past, many articles and reviews have focused on particle properties and potential use as drug delivery systems, nanoreactors, smart surface coating, etalons and sensors.[1-5,20-26]

To characterize the microgel structure, small-angle scattering on particles suspended in solvents is appropriate to characterize spatial extent and internal density, as well as local chain structure. Different modelling approaches have been applied to small-angle scattering data in the past.[9,27-29] Single-compartment microgels ("core-only") can be described in first approximation by homogeneous spheres,[30] but due to the different chemical reactivities of monomers and cross-linker, the particles have soon been recognized to have a fuzzy structure: dense inside, and progressively less dense towards the outside.[31] Different feeding strategies during synthesis have been used to control the degree of fuzziness.[32] Two different "fuzzy sphere" form factors have been proposed in the past to describe such profiles, either by convoluting with a Gaussian,[9] or by adding a polynomial decay to the homogeneous sphere.[27] The advantage of the latter is that more complex geometries can be built by adding such piece-wise parabolic functions, namely a shell can be described.[28,33,34] It has to be noted, however, that in absence of isotopic labelling, it is a priori impossible to know the origin of the monomers – core or shell – contributing to a given part of the density profile. We have recently adapted a general form-free small-angle scattering data analysis procedure based on a (reverse) Monte-Carlo (RMC) optimization, describing a priori arbitrary monomer density profiles.[35] Such form-free descriptions have the advantage that no parametrisation is necessary. Such approaches have been used in the literature to check the suitability of parametrized model form factors.[36] The monomer density profiles are automatically adapted until their scattering function corresponds to the measured one by a χ-squared minimization, and the resulting profiles have been shown to be robust with respect to simulation parameters like initial conditions. The profiles of "core-only" microgels have been compared to the models proposed in the literature, and good agreement with the (parabolic) fuzzy-sphere model was found.

"Core-only" microgels are the building block of the more complex core-shell topology.[37] The outstanding property of the core-shell microgel system is its linear change in size with the temperature between the two VPTT of the polymers.[15,16,25] It is unclear how the presence of the shell triggers this behaviour. Contrast matching via the solvent using deuteration of either shell or core monomers allows highlighting selectively only one monomer type in small-angle neutron scattering (SANS), and thus separation of the core monomer profile from the shell. Isotopic substitution induces slight changes in swelling, as shown by us recently for pNIPMAM,[14] but has no effect far from the transition temperatures.

Several core-shell microgels have been investigated over the past decade. Berndt et al. identified a core-shell structure for a microgel containing a pNIPMAM-core and a pNIPAM-shell.[28] This microgel system showed a linear change in size between the two VPTTs of the polymers, similar to the microgel system studied in this work. They explained this uncommon swelling behaviour with a restriction of the shell exerted on the core: when decreasing the temperature, at intermediate temperatures the shell impedes the swelling of the core. Zeiser et al. described this as a "corset-effect" of the "shell-corset" on the core for pNIPMAM-core pNNPAM-shell microgels,[15] in analogy with the pressure exerted by a surrounding gel matrix on microgel particles studied by Meid et al.[38]

In the present paper we report on the synthesis and structural analysis of core-shell microgel particles made of a pNIPMAM-core and a pNNPAM-shell. This system is similar to the one studied by Zeiser,[15] with the exception of the radius. The smaller size allows capturing the complete form factor in a standard SANS experiment, and thus allows a more detailed data analysis as outlined below, providing a structural understanding of the nature of the corset-effect. This has been motivated by recently published evidence from infrared spectroscopy indicating interpenetration of the core and shell polymers.[16] The present experiment has been designed to check the spatial structure of the core within a shell, using the same core polymer as Berndt et al. (pNIPMAM, VPTT $\approx$ 45 °C),[10-14,28] but a different polymer (pNNPAM, VPTT $\approx$ 23 °C)[14] for the second synthesis step, inducing a different swelling behaviour. Berndt et al. used pNIPAM as shell polymer with a VPTT of ca. 33 °C,[2,8,10,12,14] thus the linear region was reduced (covering a range from ca. 35 to 42 °C) compared to the present study. It is emphasized that the "core-only" particle studied in detail in our previous article[35] are *identical* to the ones used here, i.e. the shell synthesis has been continued with the same samples and under the same conditions as in the work by Zeiser et al.[15] We have investigated a series with different core cross-linker contents (CCC), for a temperature series (15, 30, 35, 40, and 55 °C) encompassing the transition temperatures of both polymers.

## Results and Discussion

### Swelling behaviour

After having analysed the structure of the pNIPMAM-core in our previous work,[35] we have synthesized pNIPMAM-pNNPAM core-shell microgels by polymerizing NNPAM on the *same* pNIPMAM microgels as seeds. The swelling behaviour of a pNIPMAM-pNNPAM core-shell microgel system is quite different from homopolymer microgels.[15] In Figure 1, we show the influence of the temperature on the hydrodynamic radius of this core-shell system. After the synthesis (70 °C, see exp. section), the microgels are fully collapsed at 55 °C (red circles) as shown in Figure 1. By decreasing the temperature below the VPTT of pNIPMAM of 45 °C,[10-14,28] the pNIPMAM-core would normally swell (see Fig. S1) but seems to be hindered by the pNNPAM-shell which has a VPTT of 22 °C.[11] This results in a linear increase in size from 45 °C down to 25 °C. Below the linear region, the volume phase transition of pNNPAM takes place, and at low temperatures, the whole system is swollen. The black squares in Figure 1 show the same system with deuterated D7-pNIPMAM. As the VPTT of deuterated microgels is known to shift by 3-8 K towards higher temperatures, depending on the monomer species and the level of deuteration,[14,17,18] the linear region is shortened, but the (de)swelling behaviour is qualitatively the same. The pNNPAM transition is observed to be shifted by ca. 5 K, thus in the same temperature range as reported for pNIPAM and pNIPMAM.[14,17,18] In a previous study, we investigated the deswelling and swelling behaviour of this system with different CCC by PCS in detail (Figure S10 in the SI shows the full reversibility of our microgel system. No hysteresis effects are observed in cooling/heating cycles.),[25] and we have shown that the magnitude of the slope of the linear region of the swelling curves increases with a decrease in CCC. A similar swelling behaviour was observed by Berndt et al. for pNIPMAM-core pNIPAM-shell microgels.[28] They found a much shorter linear region due to the higher VPTT of pNIPAM. Under certain conditions, two size-transition steps for the inverse system, a pNIPAM-core with a pNIPMAM-shell, were found.[39] This can be intuitively understood from the inverse geometry, which in this case of low crosslinking and small shell thickness allows the shell to swell first without being hindered by the core, apart from the interface between the two compartments. In the opposite case of high thickness and cross-linking, the shells dominate the whole swelling process.[39]

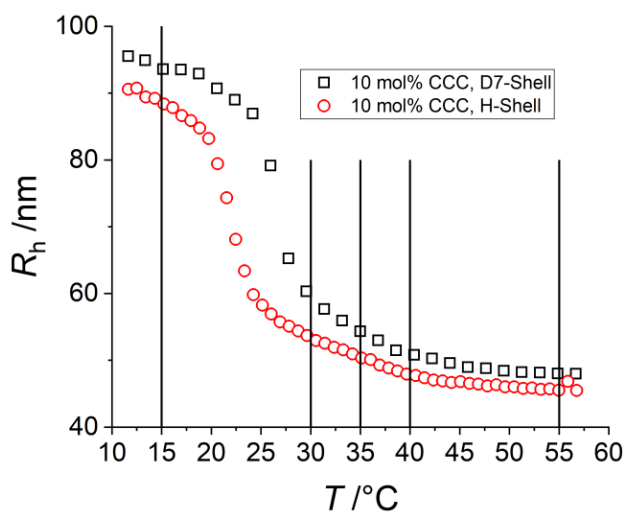

**Figure 1:** Hydrodynamic radius of core-shell particles vs. the temperature measured by PCS in H₂O. H-pNIPMAM-core (CCC = 10 mol%) with a hydrogenated (red circles) and a deuterated (black squares) pNNPAM-shell (CC = 1.9 mol%). The vertical lines indicate the temperatures where the SANS measurements were performed (15, 30, 35, 40 and 55 °C).

### H-core and H-shell system

SANS measurements have been performed to investigate the origin of the unusual (de-)swelling behaviour of pNIPMAM-pNNPAM core-shell microgels presented in Figure 1. In Figure 2a the SANS data of the non-deuterated samples have been analysed with the reverse Monte Carlo algorithm outlined in the Materials and Methods section, and the scattered intensities are well-described by our approach. For this system, the RMC algorithm does not distinguish between the two monomers because they have the same SLD (see Materials and Methods). Thus, it is a two component system with water and polymer. From a visual inspection of the data, it appears that the high-temperature microgels are smaller than at low temperature (Guinier domain shifted to the right), that the particle size is better defined (oscillation better resolved at intermediate $q$), and that the polymer chains are less visible due to reduced solvation (noisier high-$q$ data). Moreover, the data at 15 °C show the onset of a weak structure factor influence due to repulsive interactions. However, the form factor describes very well the overall decrease and the form factor intensity for $I(q{\to}0)$ is known from the data for 55 °C. It is thus not necessary to describe the relatively small perturbation (over a limited $q$-range) by the structure factor. Moreover, we have checked that our analysis is robust with respect to taking into account or not the data in this very low $q$-range (see Figure S11 in the SI). The resulting density profiles are shown in Figure 2b. It is recalled that both monomers are hydrogenated and cannot be differentiated by neutron scattering, leading to a common description of all monomers.

At high temperatures the polymer is nearly fully packed (ca. 98 %) in the centre and decreases with a high steepness to a plateau with a density of ca. 25 % from some 22 to 32 nm and a low steepness at the interface ($r \approx 50$ nm). Such a plateau was absent in pure cores, and is therefore an effect of the additional polymerization of shell monomers – it will be shown, however, that it cannot be identified with only the shell itself. For comparison, the "core-only" pNIPMAM microgel had at the same temperature a density of approximately 55 % in the centre with a high steepness of the interface and a size of some 24 nm (see corresponding results below).[35] The core-shell particle is thus denser in the centre than the "core-only" microgels, and larger in size, the latter being true at all temperatures. At intermediate temperatures (35 °C) the density is around 50 % in the centre and decreases with a low steepness at the interface. At this temperature the pNIPMAM core is in the swollen state, while the pNNPAM shell still is in the collapsed state. If we compare the monomer densities to the one in the centre of the "core-only" particles,[35] it is found that the monomer density in the centre of the core-shell particle is again higher. At low temperatures, finally, both polymers are in the swollen state and the monomer density is at about 20 % in the centre and decreases very smoothly to zero at $r \approx 60$ nm with a plateau at 10-35 nm. The density of the core-only system is ca. 15 %, thus quite similar to the core-shell particle but still lower.

While the increase in size is intuitively expected for a pNIPMAM-pNNPAM core-shell particle as compared to its own core, one may try to understand the higher density based on a compression of the shell on the core at high temperatures due to the additional polymerization. At intermediate temperatures, a restriction of the core swelling by the shell,[28,39-41] was proposed in the literature, also described as "corset-effect".[15] This model assumes the existence of a well-separated shell just outside the core, and it suggests that the shell contracts the core. The core is thus expected to be smaller, and therefore also denser, as observed for H-H core-shell microgels in the literature[27,28,39-41] and in Figure 2. Using partial deuteration, however, it will be shown below that this model is most-likely too simplified for such systems.

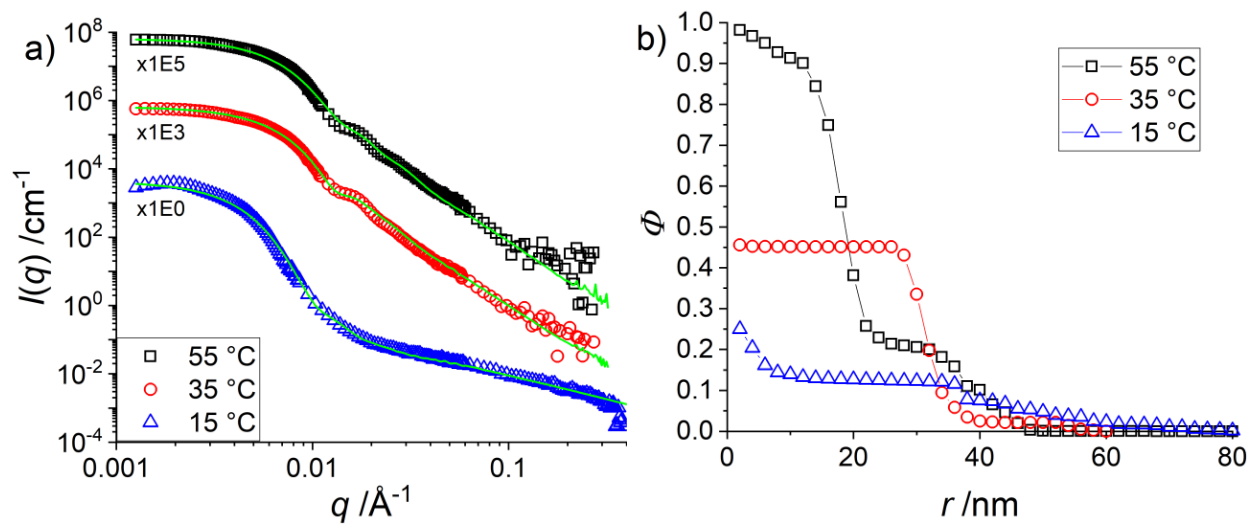

**Figure 2:** H-pNIPMAM core with H-pNNPAM shell with CCC = 10 mol% at different temperatures. **(a)** Intensity shifted experimental data of $I(q)$ vs. $q$ of SANS measurements in D₂O. The green lines indicate the multi-shell RMC fits. **(b)** Density profiles obtained with the RMC model.

In previous publications, it was shown that the degree of core crosslinking has an impact on the swelling of pNIPMAM-pNNPAM core-shell microgels.[15,25] Surprisingly, for "core-only" particles, higher crosslinking was found to lead to higher core hydration.[35] In order to study the influence of a shell on this effect of the core crosslinking, the structure of H-H core-shell microgels of various CCC (5, 10, 15 %mol) has been measured by SANS at three temperatures, 55 °C, 35 °C and 15 °C.

In Figure 3a we present the intensity-shifted SANS data and in Figure 3b the corresponding density profiles at the highest temperature (the others are shown and discussed in the SI). The intensities and the profiles are very similar at all three CCC, with a better definition of the form factor oscillation for the higher crosslinking, indicating a better defined particle with a steeper interface. For all three systems the density in the centre approaches 100 %, with a higher water content for the lowest CCC. The profile then decreases with a high steepness to a plateau at around 20 % at 20 nm. This plateau has been recognized in Figure 2b as being caused by the added shell. From 30 nm on the density decreases smoothly to the particle surface at ca. 50 nm.

As mentioned, the opposite tendency of a lower core density with higher crosslinking was found in the previous work on "core-only" particles.[35] This was explained by the lower local network flexibility with more cross-linking points, thus forcing them to accommodate water molecules. A decrease in cross-linker content leads to an increase of the degree of polymerization of the network chains between crosslinks rendering the network more flexible. The present CCC variation thus shows that the shell monomers have an impact on the swelling behaviour of the core, but it is unclear why they would compress the stiffer cores more strongly. Also, from Figure 3b the position of the shell does not seem to evolve. These results thus suggest to study where the shell monomers actually are, and how they are incorporated into the core-shell microgels to understand why they drastically affect the core swelling.

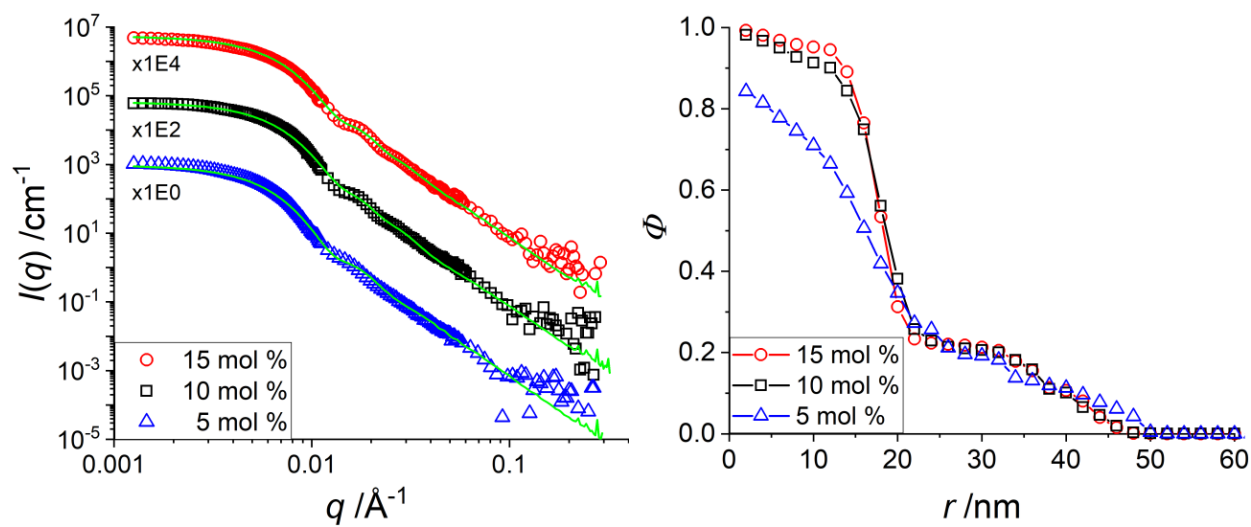

**Figure 3:** H-pNIPMAM core with H-pNNPAM shell with CCC = 5 mol% (blue triangles), CCC = 10 mol% (black squares) and CCC = 15 mol% (red circles) **(a):** $I(q)$ vs. $q$ of a SANS measurement at 55 °C in $D_2O$ **(b):** Corresponding density profiles obtained by multi-shell RMC model.

When the temperature is decreased to 35 °C and 15 °C, the density profiles are qualitatively very similar for 10 mol% and 15 mol% of core crosslinking, with a volume fraction in the centre decreasing from densely-packed at 55 °C to ca. 40 % and finally 25 %, accompanied by a large spatial extension. On the contrary, the 5 mol% CCC has a fuzzier and more temperature-insensitive structure. The difference in profile is due to the less cross-linked and thus more flexible network, and again the presence of the shell monomers induces a qualitatively different behaviour, pushing towards a denser core structure at high temperature.

**H-core profiles with matched D7-pNNPAM shell**

To investigate the structure of the core-shell microgels resolving the core monomers separately, we used the same H-pNIPMAM microgels as seeds for a polymerization with partly deuterated D7-NNPAM. We then performed SANS measurements with a $H_2O/D_2O$ ratio (12 v% $H_2O$) that matches the SLD of D7-pNNPAM (see Fig. S1) leaving only the H-pNIPMAM core visible to neutrons. The SANS intensities of both systems are shown in Figure 4a (pNIPMAM-"core-only") and 4b (core-shell system).

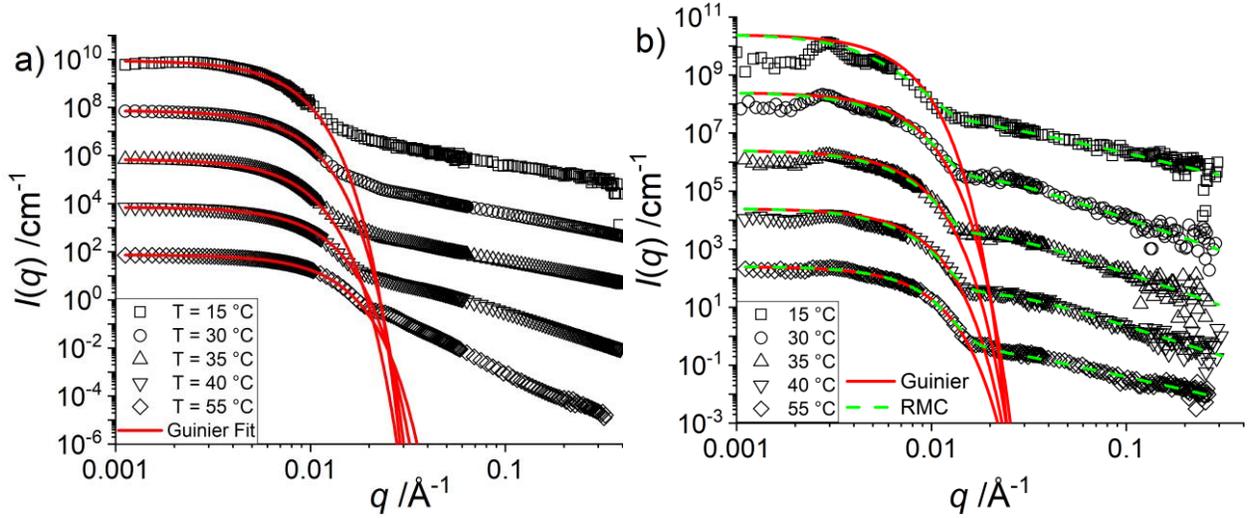

**Figure 4:** Temperature dependence of $I(q)$ vs. $q$ of SANS measurements at a concentration of 0.1 wt% and the respective Guinier fits (red continuous lines) of **(a)** A H-pNIPMAM core with a CCC of 10 mol% in $D_2O$ (exp. data adapted from Cors et al.[35] and **(b)** The same core with a matched D7-pNNPAM shell (12 v% $H_2O$), and the corresponding Guinier fits (red continuous lines), and the multi-shell RMC fits (green dashed lines).

In Figure 4b we observed the influence of a structure factor $S(q)$ for the core-shell system which does not occur for the "core-only" particles measured in Figure 4a. $S(q)$ represents the spatial correlations between the centres-of-mass of the microgels, which is caused by steric interactions. For the core-shell system, $S(q)$ contributes most for low temperatures because of the larger overall size. It progressively disappears for the higher temperatures approaching the fully collapsed state at 55 °C, as here the particles are the most compact. However, even at lower temperature, the form factor describes correctly the overall decrease, whereas the modulation induced by the structure factor remains comparatively small. Moreover, the position of the structure factor peaks can be related to the mass of each particle, via the volume conservation, assuming a simple cubic structure. It is found that this leads to a number of core monomers $N_{mono}$ which is of the same order of magnitude as the one determined from $I(q \rightarrow 0)$, see SI for further details.

As the dry volume of monomers in the core is the same in both systems, the number of core monomers and thus $I(q \rightarrow 0)$ in absence of structure is known in both systems, and given via the volume and contrast in eq. (2).[35] Together with the overall shape of the curves, this allows a description of the scattered intensity based on the model-free Guinier expression given by the equations in the experimental section, without having to quantify the exact influence of the structure factor at low $q$. The resulting low-$q$ fit is given by the red lines in Figure 5, and the corresponding Guinier radii $R$ in Table 1.

| $T$ / °C | $R$ / nm (core) | $R$ / nm (core in core-shell) |
|---|---|---|
| **55** | 27 | 36 |
| **40** | 33 | 40 |
| **35** | 42 | 43 |
| **30** | 42 | 45 |
| **15** | 47 | 51 |

**Table 1:** Temperature dependence of Guinier radii $R$ of "core-only" H-pNIPMAM microgel (CCC of 10 mol%) compared to the same core of a core-shell microgel embedded in an index-matched D7-pNNPAM shell.

The first striking observation in Table 1 is that the "core-only" microgel particles are *smaller* than the cores in the core-shell particles with matched shell at all temperatures. This is the exact opposite of the expectation of the "corset model" which predicts a compression of the core by the shell. Then, as the temperature is decreased and the particles swell, the size of the "core-only" microgel approaches the sizes of the core in the core-shell particle. The cores are thus not compressed by the shells. After this 'model-free' analysis based only on the Guinier regime, the next step is to make use of the absolute intensities, and describe the mass distribution via the volume fraction profiles in detail.

The fits resulting from the reverse Monte Carlo analysis of the spatial distribution of core monomers in the core-shell system have been superimposed to the scattered intensities in Figure 4b. The agreement between model fits and experimental intensities at intermediate and large $q$ is very good for all temperatures, whereas the oscillations induced by the structure factor at low $q$ cause deviations between the measured and the calculated intensity. In Figure 5, the results of the multi-shell RMC analysis are presented. In Figure 5a, the profiles of the core-only particles have been adapted from a previous article,[35] whereas the profiles corresponding to the scattered intensities with matched pNNPAM shell shown in Figure 4b are plotted in Figure 5b. At high temperatures (55 °C) both polymers of the microgel are collapsed, and the core of the core-shell particle is larger by some 12 nm (ca. 50 %) than its isolated counterpart. As the number of core monomers is the same – note that the cores are chemically identical –, the density of the core monomers has decreased in presence of index-matched shell monomers, as can be seen in Figure 5b (see SI for detailed plots). Moreover, the density (30 %) in the centre of the core in the core-shell particle is only almost half the density of the core-only particle (56 %), whereas it was much higher for the fully hydrogenated core-shell particles (Figure 2b). For the core-shell particle, the core density decreases with a smoothly decaying function, whereas it is constant in the core-only particle, with a sharp interface at this high temperature. As there is no reason for the core monomers at this temperature to accept more solvent, this implies that the shell monomers have penetrated and replaced water in the core during synthesis. In order to investigate the process of formation of the interpenetrated networks, we have performed PCS on the cores in presence of shell monomers, before adding the initiator (Figure S13 in the SI). It was found that these compound particles shrink slightly, which we interpret as a change in swelling of the core due to the partial replacement of water by the shell monomers. Hence, the shell does not compress the core as assumed in the corset model. Moreover, one can conclude from the smooth decay that the density of the shell monomers decreases progressively as one reaches the centre.

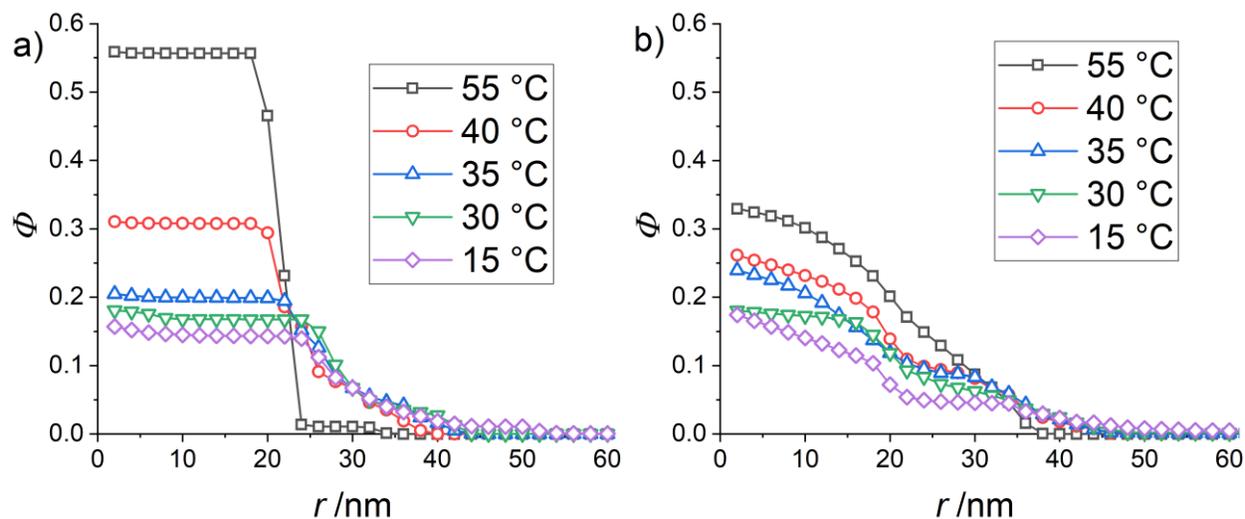

**Figure 5: (a):** Density profiles of the H-pNIPMAM-core with a CCC of 10 mol% in $D_2O$ (data adapted from Cors et al.[35] **(b)** Density profiles of the core monomers of the H-pNIPMAM-core D7-pNNPAM-shell system with the same core as in a) and an index-matched shell (CCC = 1.9 mol%).

As the temperature is decreased, the rather broad (6 K) VPT of the core (pNIPMAM) at ca. 45 °C is reached.[14] At 40 °C the core is almost fully swollen, and the difference in core radii is reduced. One may understand this as the result of swelling of the core by both water and the same amount of (still) hydrophobic shell monomers as at 55 °C. As a result, there is considerably more water in the core at 35 °C than at 55 °C, but the core monomers keep their smoothly decreasing profile in the presence of the shell, as opposed to the homogeneous core-only profiles. The average volume fractions are similar in both cores, only that with a shell the core is smoothly decaying, due to the remaining interpenetration by the shell. The maximum spatial extension of the cores seems to be roughly identical in both cases at 35 °C.

At temperatures from 35 °C to 15 °C the core is fully swollen, while the shell monomers become hydrophilic only below their VPTT of 22 °C. Thus at 15 °C, the profiles in terms of density and radii are (nearly) the same for both systems – "core-only" and core in core-shell – for these temperatures, with a core swollen not only by water, but also by shell monomers. This finding, in particular the swelling of the core by the shell polymer and the gradual decrease of the core monomer profile, would not have been possible with PCS nor by SANS without deuteration as in studies of purely hydrogenated core-shell microgels. Contrast matching combined with the multi-shell Monte Carlo analysis is necessary to focus on the structure of the cores only.

**H-pNIPMAM-core CCC variation with matched D7-pNNPAM-shell**

The peculiar linear swelling behaviour described in Figure 1 has been shown to be tunable by the core cross-linker content,[15,25] and we have studied the corresponding structural evolution above. The surprising result of higher water content with higher CCC[35] of pure "core-only" microgel particles was found to be inverted in the presence of shells (Figure 3), presumably due to the filling of the central parts of the particles by interpenetrating shell polymer. However, due to the absence of measurable contrast between both monomers in these experiments, it was not possible to isolate the spatial distribution of one of them, e.g. the core monomers. Here, the core of the core-shell particles is again the same as the previously studied "core-only" system,[35] and addition of a deuterated shell allows observing the core for different CCC. In Figure 6, the scattered intensities and the corresponding radial density profiles of the core monomers are shown for $T$ = 55 °C, for three CCC (5 mol%, 10 mol%, and 15 mol%). As a first result, the size of the core with an index-matched shell is larger for all CCC as the "core-only" systems (see Figure S5): the monomer density reaches zero at ca. 40 nm to 50 nm, while the "core-only" particles only extends to a radius of approximately 25 nm to 30 nm.[35] Moreover, the steepness of the core interface is lower for all CCC compared to the "core-only" surface. These observations are compatible with the swelling of the core by the shell polymer. The steepness of the lowest CCC is higher compared to 10 and 15 mol%. This finding is in good agreement with the "core-only" systems, where the higher flexibility of the loosely cross-linked chains was invoked. The low steepness at 10 and 15 mol% is comparable to the interface of a swollen particle, indicating again a high swelling of the core by the shell monomers. Finally, the core monomer density is lower than the one of the "core-only" particles, due to the filling of the core by interpenetrating shell monomers. The slight dependence in Figure 6b of the core density in the centre on CCC is less pronounced but opposed to our previous result.[35] As in Figure 3b, this demonstrates again that the filling of the core by shell monomers provides an additional degree of freedom compensating for the higher rigidity of the more cross-linked samples.

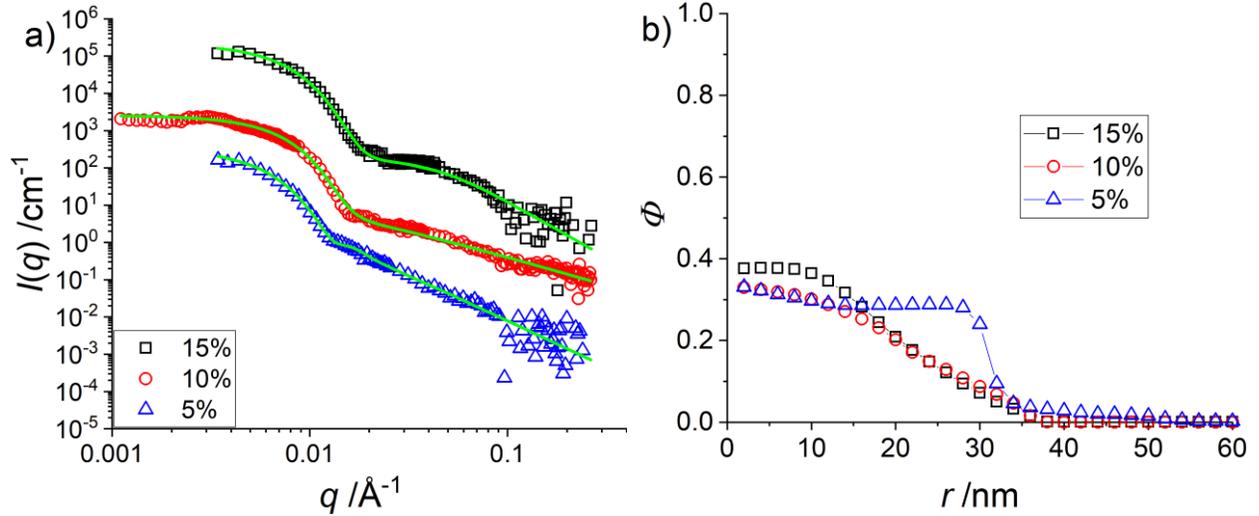

**Figure 6:** (a) $I(q)$ vs. $q$ from SANS measurements of three H-pNIPMAM-core D7-pNNPAM-shell microgels with different CCC as indicated in the legend at $T = 55$ °C, under matching conditions of the shell. (b) Density profiles of core monomers determined by multi-shell RMC model.

**Evolution of density profiles with temperature**

The above comparisons between cores with deuterated (Figure 5b) and hydrogenated (Figure 2b) shells suggest that the remaining space is filled by shell monomers. By combining the corresponding density profiles ($\Phi_{core}$ and $\Phi_{coreshell}$ for the profiles of the core measured with deuteration, and for all monomers measured in the fully hydrogenated system, respectively), one may deduce the a priori unknown density profile of the shell $\Phi_{shell}$, for each temperature:

$$\Phi_{core}(r) + \Phi_{shell}(r) = \Phi_{coreshell}(r) \qquad (1)$$

Moreover, one may also infer the volume fraction profile of the solvent from the core-shell measurement:

$$\Phi_{water}(r) = 1 - \Phi_{coreshell}(r) \qquad (2)$$

The resulting volume fraction profiles for core-shell microgels with CCC = 10 mol% are shown in Figure 7. Starting with the highest temperature, the shell monomers are seen to strongly interpenetrate the core. Moreover, just outside the core they contribute to the formation of the plateau observed in the measurements with hydrogenated shell. In Figure 7b, below the transition temperature of the core, the water content is found to increase to a constant level in the particle, which is seen to be made of a shell of progressively increasing density, at the expense of the decaying core monomer profile. At the lowest temperature (Figure 7c), finally, the same type of progressive replacement of the core by the shell monomers is found, in presence of a much higher water content.

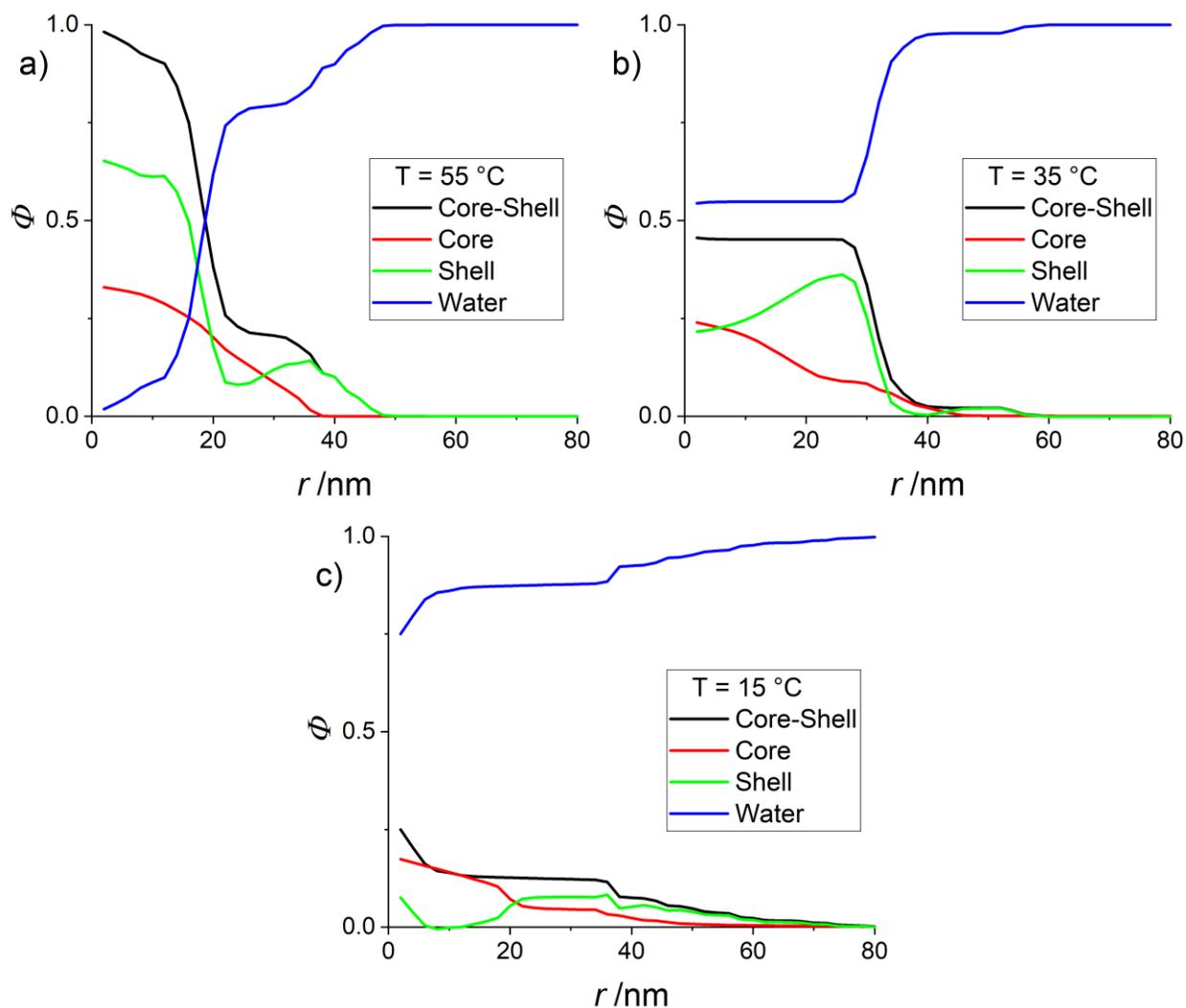

**Figure 7:** Density profiles of core and shell monomers, and water for H-pNIPMAM/D7-pNNPAM core-shell microgels (CCC = 10 %mol) in $H_2O/D_2O$ under matching conditions of the shell. (a) $T$ = 55 °C (b) $T$ = 35 °C (c) $T$ = 15 °C.

One has to keep in mind, however, that $\Phi_{core}(r)$ and $\Phi_{coreshell}(r)$ result from measurements on two different samples, namely hydrogenated and deuterated ones. The present analysis is thus speculative, and holds only if the particles are indeed very similar. Given the very similar radii ($R_h$=48 nm (H-shell) $R_h$ = 45.5 (D7-shell)), this hope seems reasonable, but only a third measurement, highlighting the shell and matching the core-monomers, can definitely answer this question.

# Conclusion

The origin of the peculiar linear swelling behaviour of pNIPMAM-core and pNNPAM-shell core-shell microgel particles has been investigated by studying the structure of the microgels as a function of temperature, with a special focus on the shape of the core described by its volume fraction profile. The latter could be accessed by performing SANS experiments on microgels with selectively deuterated pNNPAM-shells, and matching them in appropriate $H_2O/D_2O$ mixtures. The density profile was obtained by applying a form-free multi-shell model and seeking for appropriate solutions using a reverse Monte Carlo approach, which goes beyond simple parametrisation and directly yields the detailed monomer density distribution inside the particles. As a reference case, otherwise identical purely hydrogenated core-shell particles have also been studied. Using the conservation of volume and the hypothesis of identical H/H and H/D-particles, they allow the determination of all relevant profiles, namely of water, core monomers, and shell monomers.

In parallel, a model-free approach based on the analysis of the Guinier regime has been employed, comparing pure "core-only" particles with the cores embedded in matched shells. All these results, including cross-linker variations, point towards the same conclusion: the cores of such core-shell microgels are not compressed, but interpenetrated and swollen by the shell monomers. They thus reach a greater spatial extent, and a lower core-monomer density, which smoothly decays as the shell density increases towards the outer particle surface. The core and the shell are strongly interpenetrated, as already suggested by IR measurements. Given the different phase transition temperatures, the core swells first under decreasing temperatures, while it is hindered by its interpenetration with the still hydrophobic shell monomers, apparently resulting in a linear swelling. To summarize, our SANS results combined with reverse Monte Carlo allowed the precise determination of the degree of penetration of the shell down to the centre of the core.

As already mentioned, the shell density profile has been deduced in an indirect way, using a combination of core and (total) core-shell densities determined for different samples of very similar structure. Only an independent measurement matching the core and evidencing the shell will show if the last assumption of identical samples regarding all but deuteration holds. Such work is currently in progress. It is hoped that the results and methods proposed in this article will be useful for studies of similar systems, and in particular allow a deeper understanding and thus possible tuning of the linear swelling behaviour.

## Materials and Methods

**Synthesis:** The synthesis of poly(*N*-isopropylmethacrylamide) (pNIPMAM) "core-only" microgels used as seed for core-shell particles in this article was described in detail elsewhere.[25,42] In short, we synthesized hydrogenated (H12-pNIPMAM, Sigma-Aldrich, St. Louis, USA) and deuterated (D12-pNIPMAM, Polymer Source Inc., Montreal, Canada) microgels following a precipitation polymerization protocol with *N*,*N*'-methylenebis(acrylamide) (BIS, Sigma-Aldrich, St. Louis, USA, 99%) as cross-linker. We synthesized pNIPMAM microgels with three different CCC: 5, 10 and 15 mol%. We then used these particles as seed for a second polymerization step with *N*-*n*-propylacrylamide (NNPAM) as monomer. The second polymerization step was also done with hydrated (NNPAM) and partly deuterated (D7-NNPAM) monomer and a cross-linker content (CC) of 1.9 mol%. The (deuterated) NNPAM was synthesized following a Schotten-Baumann reaction as described by Hirano et al.[43] using acryloyl chloride (97 %, Aldrich, U.S.A.), triethylamine (99 %, Grüssing, Germany), propylamine (99 %, Fluka, U.S.A.) and methylene chloride (purity ≥99.8 %, VWR International, Fontenay-sous-Bois, France) as solvent. For the deuterated NNPAM 2 g of *n*-propyl-D7-amine (CDN isotopes, Canada) was added together with 2.67 g of triethyleamin to 10 mL dichloromethane. A solution of 2.72 g acryloyl chloride in 5 mL dichloromethane was added dropwise while the solution was cooled with an ice bath. The resulting solution was stirred at room temperature for 24 hours, washed with a 10 wt% sodium hydrogen carbonate solution and dried with magnesium sulfate. The solvent was removed by rotary evaporation. The product was obtained with a yield of ca. 35 % by vacuum distillation (10 mbar, 125 °C oil bath temperature).

**Photon correlation spectroscopy (PCS):** Temperature-dependent dynamic light scattering measurements were done at a fixed scattering angle of 60° (angular dependent measurements were already reported in a previous article.[35] The experimental setup was partially home-made with a He-Ne laser (HNL210L-EC, 632.8 nm, Thorlabs, Newton, USA) a detector (SO-SIPD, ALV GmbH, Langen, Germany) and a digital multiple tau hardware correlator (ALV-6010, ALV GmbH, Langen, Germany). The sample placed in a decalin bath to match the refractive index of the goniometer windows and the cuvette. The temperature was controlled using a refrigerated bath (Haake C25P, Thermo Fisher Scientific, Waltham, USA) equipped with a controller (Phoenix II, Thermo Fisher Scientific, Waltham, USA). Results are reported in terms of the hydrodynamic radius $R_h$ as a function of temperature.

**SANS measurements:** The SANS measurements were done at KWS-1[44] (JCNS at MLZ, Garching, Germany) and PA20[45] (LLB, Saclay, France). To cover the *q*-range of 0.001 Å⁻¹ to 0.3 Å⁻¹ three different configurations were used on both instruments. On the KWS-1 we used: 1 m and $\lambda$ =5 Å, 8 m and 5 Å, and 20 m and 12 Å. The configurations on PA20 were: 1.5 m and 4 Å, 8 m and 6 Å, and 20 m and 6 Å. The normalization (detector electronic noise, water or Plexiglas, empty cell) of the data was done with QtiKWS (JCNS, Germany) for the data from KWS-1 and with Pasinet software (LLB, Saclay, France) for the data from PA20 to receive scattering data in absolute units.

The scattering length densities (SLD) of the microgel components were determined by contrast variation. In the past we determined the SLD of pNIPMAM to be $1.0 \cdot 10^{10}$ cm$^{-2}$.[35] The contrast variation in Figure S1 of the SI reveals a SLD of $5.54 \cdot 10^{10}$ cm$^{-2}$ for D7-pNNPAM, and the contrast variation of H-pNNPAM in Figure S2 reveals a SLD of $1.05 \cdot 10^{10}$ cm$^{-2}$. The SLD of the solvents were estimated assuming a density of 1.107 g/cm$^3$ for D$_2$O (SLD = $6.38 \cdot 10^{10}$ cm$^{-2}$) and of 1.000 g/cm$^3$ for H$_2$O (SLD = $-0.56 \cdot 10^{10}$ cm$^{-2}$).

**Analysis of SANS data by multi-shell reverse Monte-Carlo Simulations:** A form-free multi-shell model with Monte-Carlo optimization of fit agreement with SANS data has been presented recently and applied to simple "core-only" microgels.[35] The model is based on a coarse-grained radial monomer volume fraction profile which is modified by the algorithm until its scattering prediction agrees best with the experimental intensity, while taking into account constraints, and including polydispersity in size (20 % for the particles in this article).[35] Our RMC Algorithm calculates $\chi^2$ of the experimental data and the intensity curve of the current profile after each Monte Carlo step of monomer motion. The algorithm then rejects or accepts the MC step, in order to minimize $\chi^2$ by simulated annealing. Each calculated density profile is thus the outcome of a $\chi$-squared minimization, including averaging once an acceptable $\chi$-squared is found (See Table S2 in the SI for the values). However, residual structure factor effects are not taken into account at present. As a rule-of-thumb, we would say that profiles are trustworthy within some 5 %, as we found when running different fits on the same data. Among the constraints, a smoothly decaying profile was sought, forbidding in particular empty shells in the middle of the microgel particles. The approach assumes a spherical symmetry of the particle represented by several concentric shells filled with water and monomer, and includes an ad-hoc addition of chain scattering based on the generalized coil.[46] However, it is not a simple parametrisation like the usual form factor fits and gives much more detailed information in terms of the monomer density distribution. Details of the modelling approach are described elsewhere.[35] In the present work we exploit this approach combining it with the possibilities arising from contrast variation by means of partial or total deuteration of one of the monomers. This strategy allows to describe the spatial distribution of core monomers in a core-shell microgel with a contrast-matched shell which thus does not contribute to the scattering and which does not need to be described in the model. Modelling results for the core surrounded by the shell are then compared to previously determined monomer density profiles of "core-only" microgels,[35] and the effect of the shell on the core can thus be determined. For a 'model-free' comparison, we also use the Guinier expression for the low-$q$ intensity describing finite-sized objects of radius of gyration $R_g$:[47]

$$I(q) = I_0 \exp(-q^2 R_g^2/3) \qquad (3)$$

where $I_0$ is a prefactor describing $I(q{\to}0)$ and given by the product of the dry volume $V$ of material of an object contributing to the scattered intensity, of the volume fraction $\Phi$ of this material, and the square of the contrast $\Delta\rho^2$. In case of a binary polymer-solvent system, this prefactor reads:

$$I_0 = \Phi \, \Delta\rho^2 \, V \qquad (4)$$

It is often convenient to express the radius of gyration as the radius $R$ of the equivalent monodisperse homogeneous sphere (called the 'Guinier' radius in this article):

$$R^2 = 5/3 \, R_g^2 \qquad (5)$$

## References

1.    Murray, M. & Snowden, M. The preparation, characterisation and applications of colloidal microgels. *Advances in Colloid and Interface Science* **54**, 73–91 (1995).

2.    Saunders, B. R. & Vincent, B. Microgel particles as model colloids: theory, properties and applications. *Advances in Colloid and Interface Science* **80**, 1–25 (1999).

3.    Pelton, R. Temperature-sensitive aqueous microgels. *Advances in Colloid and Interface Science* **85**, 1–33 (2000).

4.    Saunders, B. R. *et al.* Microgels: From responsive polymer colloids to biomaterials. *Advances in Colloid and Interface Science* **147-148**, 251–262 (2009).

5.    Richtering, W. & Saunders, B. R. Gel architectures and their complexity. *Soft Matter* **10**, 3695–3702 (2014).

6.    Karg, M. *et al.* Nanogels and microgels: From model colloids to applications, recent developments, and future trends. *Langmuir* **35**, 6231–6255 (2019).

7.    Scherzinger, C. *et al.* Cononsolvency of mono- and di-alkyl n-substituted poly(acrylamide)s and poly(vinyl caprolactam). *Polymer* **62**, 50–59 (2015).

8.    Hirokawa, Y. & Tanaka, T. Volume phase transition in a nonionic gel. *The Journal of Chemical Physics* **81**, 6379–6380 (1984).

9.    Stieger, M., Richtering, W., Pedersen, J. S. & Lindner, P. Small-angle neutron scattering study of structural changes in temperature sensitive microgel colloids. *The Journal of Chemical Physics* **120**, 6197–6206 (2004).

10.     Balaceanu, A., Demco, D. E., Möller, M. & Pich, A. Heterogeneous morphology of random copolymer microgels as reflected in temperature-induced volume transition and 1H high-resolution transverse relaxation NMR. *Macromolecular Chemistry and Physics* **212**, 2467–2477 (2011).

11.     Wedel, B., Zeiser, M. & Hellweg, T. Non NIPAM based smart microgels: Systematic variation of the volume phase transition temperature by copolymerization. *Zeitschrift für Physikalische Chemie* **226**, 737–748 (2012).

12.     Wu, Y., Wiese, S., Balaceanu, A., Richtering, W. & Pich, A. Behavior of temperature-responsive copolymer microgels at the oil/water interface. *Langmuir* **30**, 7660–7669 (2014).

13.     Crassous, J. J., Mihut, A. M., Månsson, L. K. & Schurtenberger, P. Anisotropic responsive microgels with tuneable shape and interactions. *Nanoscale* **7**, 15971–15982 (2015).

14.     Cors, M., Wiehemeier, L., Oberdisse, J. & Hellweg, T. Deuteration-induced volume phase transition temperature shift of PNIPMAM microgels. *Polymers* **11**, 620 (2019).

15.     Zeiser, M., Freudensprung, I. & Hellweg, T. Linearly thermoresponsive coreshell microgels: Towards a new class of nanoactuators. *Polymer* **53**, 6096–6101 (2012).

16.     Wiehemeier, L. *et al.* Swelling behaviour of coreshell microgels in $H_2O$, analysed by temperature-dependent FTIR spectroscopy. *Physical Chemistry Chemical Physics* **21**, 572–580 (2019).

17.     Mohanty, P. S. *et al.* Interpenetration of polymeric microgels at ultrahigh densities. *Scientific Reports* **7** (2017).

18.     Nöjd, S. *et al.* Deswelling behaviour of ionic microgel particles from low to ultra-high densities. *Soft Matter* **14**, 4150–4159 (2018).

19.     Kratz, K., Hellweg, T. & Eimer, W. Structural changes in PNIPAM microgel particles as seen by SANS, DLS, and EM techniques. *Polymer* **42**, 6631–6639 (2001).

20.     Lu, Y. *et al.* Thermosensitive core-shell microgel as a nanoreactor for catalytic active metal nanoparticles. *Journal of Materials Chemistry* **19**, 3955 (2009).

21.     Sorrell, C. D., Carter, M. C. D. & Serpe, M. J. A paint-on protocol for the facile assembly of uniform microgel coatings for color tunable etalon fabrication. *ACS Applied Materials & Interfaces* **3**, 1140–1147 (2011).

22.     Li, X. & Serpe, M. J. Understanding and controlling the self-folding behavior of poly (*N*-isopropylacrylamide) microgel-based devices. *Advanced Functional Materials* **24**, 4119–4126 (2014).

23.     Zhang, Q. M., Berg, D., Mugo, S. M. & Serpe, M. J. Lipase-modified pH-responsive microgel-based optical device for triglyceride sensing. *Chemical Communications* **51**, 9726–9728 (2015).

24.     Wellert, S., Richter, M., Hellweg, T., von Klitzing, R. & Hertle, Y. Responsive microgels at surfaces and interfaces. *Zeitschrift für Physikalische Chemie* **229**, 1225–1250 (2015).

25.     Cors, M. *et al.* Coreshell microgel-based surface coatings with linear thermoresponse. *Langmuir* **33**, 6804–6811 (2017).

26.     Wrede, O. *et al.* Volume phase transition kinetics of smart *N*-n-propylacrylamide microgels studied by time-resolved pressure jump small angle neutron scattering. *Scientific Reports* **8** (2018).

27.     Berndt, I., Pedersen, J. S. & Richtering, W. Structure of multiresponsive intelligent core-shell microgels. *Journal of the American Chemical Society* **127**, 9372–9373 (2005).

28.     Berndt, I., Pedersen, J. S. & Richtering, W. Temperature-sensitive coreshell microgel particles with dense shell. *Angewandte Chemie International Edition* **45**, 1737–1741 (2006).

29.     Boon, N. & Schurtenberger, P. Swelling of micro-hydrogels with a crosslinker gradient. *Physical Chemistry Chemical Physics* **19**, 23740–23746 (2017).

30.     Pusey, P. N. Introduction to Scattering Experiments. In Lindner, T. Z. P. (ed.) *Neutron, X-Rays and Light. Scattering Methods Applied to Soft Condensed Matter*, chap. Introduction to Scattering Experiments, 3–22 (Elsevier Science B. V., 2002).

31.     Wu, X., Pelton, R. H., Hamielec, A. E., Woods, D. R. & McPhee, W. The kinetics of poly(*N*-isopropylacrylamide) microgel latex formation. *Colloid & Polymer Science* **272**, 467–477 (1994).

32.     Witte, J. *et al.* A comparison of the network structure and inner dynamics of homogeneously and heterogeneously crosslinked PNIPAM microgels with high crosslinker content. *Soft Matter* **15**, 1053–1064 (2019).

33.     Berndt, I., Pedersen, J. S., Lindner, P. & Richtering, W. Influence of shell thickness and cross-link density on the structure of temperature-sensitive poly-*N*-isopropylacrylamide-poly-*N*-isopropylmethacrylamide core-shell microgels investigated by small-angle neutron scattering. *Langmuir* **22**, 459–468 (2006).

34.     Berndt, I., Popescu, C., Wortmann, F.-J. & Richtering, W. Mechanics versus thermodynamics: Swelling in multiple-temperature-sensitive coreshell microgels. *Angewandte Chemie International Edition* **45**, 1081–1085 (2006).

35.     Cors, M. *et al.* Determination of internal density profiles of smart acrylamide-based microgels by small-angle neutron scattering: A multishell reverse monte carlo approach. *Langmuir* **34**, 15403–15415 (2018).

36.     Virtanen, O. L. J., Mourran, A., Pinard, P. T. & Richtering, W. Persulfate initiated ultra-low cross-linked poly(*N*-isopropylacrylamide) microgels possess an unusual inverted cross-linking structure. *Soft Matter* **12**, 3919–3928 (2016).

37.     Hellweg, T. Responsive core-shell microgels: Synthesis, characterization, and possible applications. *Journal of Polymer Science Part B: Polymer Physics* **51**, 1073–1083 (2013).

38.     Meid, J., Friedrich, T., Tieke, B., Lindner, P. & Richtering, W. Composite hydrogels with temperature sensitive heterogeneities: influence of gel matrix on the volume phase transition of embedded poly-(*N*-isopropylacrylamide) microgels. *Phys. Chem. Chem. Phys.* **13**, 3039–3047 (2011).

39.     Berndt, I. & Richtering, W. Doubly temperature sensitive core-shell microgels. *Macromolecules* **36**, 8780–8785 (2003).

40.     Jones, C. D. & Lyon, L. A. Shell-restricted swelling and core compression in poly(*N*-isopropylacrylamide) core-shell microgels. *Macromolecules* **36**, 1988–1993 (2003).

41.     Jones, C. D. & Lyon, L. A. Dependence of shell thickness on core compression in acrylic acid modified poly(*N*-isopropylacrylamide) core/shell microgels. *Langmuir* **19**, 4544–4547 (2003).

42.    Pelton, R. & Chibante, P. Preparation of aqueous latices with *N*-isopropylacrylamide. *Colloids and Surfaces* **20**, 247–256 (1986).

43.    Hirano, T. *et al.* Hydrogen-bond-assisted syndiotactic-specific radical polymerizations of *N*-alkylacrylamides: The effect of the *N*-substituents on the stereospecificities and unusual large hysteresis in the phase-transition behavior of aqueous solution of syndiotactic poly(*N*-n-propylacrylamide). *Journal of Polymer Science Part A: Polymer Chemistry* **46**, 4575–4583 (2008).

44.    Feoktystov, A. V. *et al.* KWS-1 high-resolution small-angle neutron scattering instrument at JCNS: current state. *Journal of Applied Crystallography* **48**, 61–70 (2015).

45.    Chaboussant, G., Désert, S., Lavie, P. & Brûlet, A. PA20 : A new SANS and GISANS project for soft matter, materials and magnetism. *Journal of Physics: Conference Series* **340**, 012002 (2012).

46.    Hammouda, B. Small-angle scattering from branched polymers. *Macromolecular Theory and Simulations* **21**, 372–381 (2012).

47.    Lindner, T. Z. P. (ed.) *Neutron, X-Rays and Light. Scattering Methods Applied to Soft Condensed Matter* (Elsevier Science B. V., 2002).

## Figure Legend

**Figure 1:** Hydrodynamic radius of core-shell particles vs. the temperature measured by PCS in $H_2O$. H-pNIPMAM-core (CCC = 10 mol%) with a hydrogenated (red circles) and a deuterated (black squares) pNNPAM-shell (CC = 1.9 mol%). The vertical lines indicate the temperatures where the SANS measurements were performed (15, 30, 35, 40 and 55 °C).

**Figure 2:** H-pNIPMAM core with H-pNNPAM shell with CCC = 10 mol% at different temperatures. **(a)** Intensity shifted experimental data of I(q) vs. *q* of SANS measurements in $D_2O$. The green lines indicate the multi-shell RMC fits. **(b)** Density profiles obtained with the RMC model.

**Figure 3:** H-pNIPMAM core with H-pNNPAM shell with CCC = 5 mol% (blue triangles), CCC = 10 mol% (black squares) and CCC = 15 mol% (red circles) **(a):** *I(q)* vs. *q* of a SANS measurement at 55 °C in $D_2O$ **(b):** Corresponding density profiles obtained by multi-shell RMC model.

**Figure 4:** Temperature dependence of *I(q)* vs. *q* of SANS measurements at a concentration of 0.1 wt% and the respective Guinier fits (red continuous lines) of **(a)** A H-pNIPMAM core with a CCC of 10 mol% in $D_2O$ (exp. data adapted from Cors et al.[35] and **(b)** The same core with a matched D7-pNNPAM shell (12 v% $H_2O$), and the corresponding Guinier fits (red continuous lines), and the multi-shell RMC fits (green dashed lines).

**Figure 5: (a):** Density profiles of the H-pNIPMAM-core with a CCC of 10 mol% in $D_2O$ (data adapted from Cors et al.[35] **(b)** Density profiles of the core monomers of the H-pNIPMAM-core D7-pNNPAM-shell system with the same core as in a) and an index-matched shell (CCC = 1.9 mol%).

**Figure 6: (a)** *I(q)* vs. *q* from SANS measurements of three H-pNIPMAM-core D7-pNNPAM-shell microgels with different CCC as indicated in the legend at *T* = 55 °C, under matching conditions of the shell. **(b)** Density profiles of core monomers determined by multi-shell RMC model.

**Figure 7:** Density profiles of core and shell monomers, and water for H-pNIPMAM/D7-pNNPAM core-shell microgels (CCC = 10 %mol) in $H_2O$/$D_2O$ under matching conditions of the shell. (a) $T$ = 55 °C (b) $T$ = 35 °C (c) $T$ = 15 °C.

## Table Legend

**Table 1:** Temperature dependence of Guinier radii $R$ of "core-only" H-pNIPMAM microgel (CCC of 10 mol%) compared to the same core of a core-shell microgel embedded in an index-matched D7-pNNPAM shell.

## Acknowledgements

The authors thank the ANF and DFG for the support with the joint CoreShellGel project, Grant ANR-14-CE35-0008-01 of the French Agence Nationale de la Recherche, and Grant HE2995/5-1 by Deutsche Forschungsgemeinschaft. We acknowledge support for the Article Processing Charge by the Deutsche Forschungsgemeinschaft and the Open Access Publication Fund of Bielefeld University.

## Authors contributions statement

J.O. and T.H. conceived the research project and acquired the funding. M. C. conceived and conducted the PCS experiments and analysed the results. M. C. and J. O. conceived the SANS experiments. M. C., L. W., A. F., and F. C. conducted the SANS experiments. M. C. and J. O. analysed the SANS results. O. W. synthesized the monomer D7-NNPAM. All authors discussed the experimental results. The main part of the manuscript was written by M. C. and J. O and corrected by T.H. All authors reviewed the manuscript.

## Additional information

**Competing Interests:** The authors declare no competing interests.